\documentstyle[preprint,aps]{revtex}

\begin{document}

\newcommand{\beq}{\begin{equation}}
\newcommand{\eeq}{\end{equation}}
\newcommand{\bea}{\begin{eqnarray}}
\newcommand{\eea}{\end{eqnarray}}
\newcommand{\rv}{\rangle}
\newcommand{\lv}{\langle}

\draft
\tighten
\preprint{\vbox{
\hbox{IJS-TP-96/18}
}}

\title{SYMMETRY RESTORATION IN HOT SUSY
\thanks{Based on the talk by G. Senjanovi\'c at SUSY96, University 
of Maryland, 1996.}}
\author{ Borut Bajc \thanks{ borut.bajc@ijs.si}}
\address{International Center for Theoretical Physics,
34100 Trieste, Italy, {\rm and} \newline J. Stefan Institute, 1001 
Ljubljana, Slovenia   } 
\author{Goran Senjanovi\'c \thanks{ goran@ictp.trieste.it}}
\address{International Center for Theoretical Physics,
34100 Trieste, Italy } 

\maketitle

\begin{abstract}

It is by now well known that symmetries may be broken at high 
temperature. However,
in renormalizable supersymmetric theories any internal symmetry gets 
always restored. 
In nonrenormalizable theories the situation is far less simple. 
We review here some recent work which seems to indicate 
that renormalizability is  not essential for the 
restoration of internal symmetries in supersymmetry. 

\end{abstract}

\newpage

\section{Introduction}
\label{intr}

One of the central issues in high temperature field theory is the 
question of what happens to the symmetries of the Lagrangian which may 
(or may not) be spontaneously broken.
 In spite of one's intuitive expectations on symmetry 
restoration \cite{kl72}, based on daily experience and proven correct
 in the simplest field theory systems \cite{kl72,w74,dj74}, one can easily 
find examples with 
symmetries  broken at high temperature \cite{w74,ms79}. This is an 
important issue, due to its possible role in the production of 
topological defects in the early universe. Symmetry nonrestoration at 
high temperature may provide a way out of both the domain wall 
\cite{ds95,dms96} and the monopole problem \cite{lp80,sss85,dms95,bl95}. 

A simple example of broken symmetry at high temperature is provided by 
supersymmetry. Due to the different boundary condition for bosons and 
fermions in thermal field theory, supersymmetry is automatically broken 
at any non-zero temperature. However, for the issue of topological 
defects, one would like to know what happens to internal symmetries 
in the context of supersymmetries. This question is nontrivial due to the 
highly constrained structure of SUSY models. It has been addressed 
carefully more than ten years ago \cite{m84}: in contrast 
with the non-supersymmetric case, it was shown that supersymmetry 
necessarily implies restoration of  internal symmetries at high temperature.
At least, this is what happens in renormalizable theories.

Recently, this conclusion was questioned by Dvali and Tamvakis 
\cite{dt96} precisely by resorting to non-renormalizable potentials. They 
present an explicit example in which the inclusion of a quartic term in 
the superpotential allows apparently for  non vanishing vevs at high 
temperature. Stimulated by their interesting suggestion,
with A. Melfo  we have analyzed 
carefully their example, arriving however to the opposite conclusion 
\cite{bms96}. We review here both the work of Dvali and Tamvakis 
and our own, and try to explain why there may be a general problem with 
the idea of Ref. \cite{dt96}. More precisely, we will see that this 
program strictly speaking is independent of SUSY, and discuss it in its 
own merit. Thus in the next section we present a general  prototype 
example (independent of SUSY) along the lines of Ref. \cite{dt96}
 and show why it cannot lead to symmetry nonrestoration. In section 3 we 
apply this  to the simple case of a supersymmetric model with a discrete 
symmetry.

In section 4 we make some remarks on the validity of perturbation theory 
applied to a general nonrenormalizable potential, and to the SUSY 
example of section 3. Finally, in the last section we offer some 
thoughts on the generality of our results. We study a potential 
counterexample to symmetry restoration based on derivative 
couplings, which at the first glance could lead to negative 
$T$-dependent mass squared for the scalar field. However, in 
the leading limit studied throughout this work we find it 
not working. Thus the challenge of proving our results in 
general SUSY theories of finding a counterexample to symmetry 
restoration at high $T$ still remains.

\section{A prototype example}
\label{ai}

The idea of Dvali and Tamvakis can be exemplified nicely on a simple
example of a real scalar field  with a symmetry {\bf D:}$\phi\to -\phi$

\beq  
V = {\mu^2 \over 2} \phi^2 - {\epsilon \over 4} \phi^4 + 
{\phi^6 \over 6 M^2}\;,
\label{scalarpot}
\eeq

\noindent
where $\epsilon$ is very small, in order for the nonrenormalizable 
interaction to play an important role: $\epsilon \ll 1$ (Dvali and 
Tamvakis take it to be of order $\mu/M$), $M$ is the large scale, 
imagined to be a GUT or a Planck scale, and $\mu \ll M$. Assuming $M^2 
>0$, the above potential is bounded even for $\epsilon > 0$.

A naive one loop expression for the high temperature potential would then be 

\bea
\Delta V_T({\rm 1-loop})&=& {1 \over 24} T^2 
{\partial^2 V \over \partial \phi^2}
 \nonumber \\ 
&=& {1 \over 24} T^2  \left[ - 3 \epsilon \phi^2 + 5 {\phi^4 \over M^2}
\right]\;,
\label{deltave}
\eea

\noindent where $T^2 \gg \mu M$, but of course $T \ll M$. Notice the well
 known fact that the sign of $\epsilon$ defines the sign of the temperature
 induced one-loop mass term (which in turn determines the pattern of
 symmetry breaking).

At sufficiently high T, it is clear that the expression (\ref{deltave})
 implies a nonvanishing vev for $\phi$:

\beq
\langle \phi \rangle^2 \simeq \epsilon M^2
\label{vev}
\eeq

\noindent as long as $\epsilon >0$. In the Dvali-Tamvakis case,  when
 $\epsilon \simeq \mu/M$, $\langle \phi \rangle^2 \simeq \mu M$ and $T^2 
\gg \mu M$ in order to justify  a high T limit for $\phi$. It is easy to 
see that then a mass term (if there at all) of the order $T^4 \phi^2/M^2$ 
would dominate  the $\epsilon T^2 \phi^2 \simeq (\mu/M) T^2 \phi^2$ one. 

In a paper with Melfo \cite{bms96}, we found out that at the two-loop level
 such term is generated with a positive coefficient, more precisely we got

\beq
\Delta V_T({\rm 2-loops}) = {5 T^4\over 96 M^2 } \phi^2\;.
\label{Tpot2l}
\eeq

This implies clearly that for $T^2 \gg \epsilon M^2 $ or $T^2  \gg \mu M$ the
 symmetry is necessarily restored.

A comment is noteworthy here. The above result should not be viewed as 
the breakdown of perturbation theory, as one may think at first sight. 
After all we are saying that the two-loop effect is larger than the one-loop 
one. However, this is a common situation in theories with more than one 
coupling. The prototype example of such a situation is the Coleman-Weinberg
 \cite{cw73}  
one-loop effective potential that may dominate the tree-level potential. 

The contribution (\ref{Tpot2l}) has its origin in the nonrenormalizable 
coupling $\phi^6/M^2$ in (\ref{scalarpot}) and it appears for the first 
time at the two-loops level. In this sense, the one-loop approximation 
fails, but not the perturbation theory itself, for after the inclusion 
of (\ref{Tpot2l}), the higher-loop effects are down by $\phi/M$, $T/M$ 
or $\epsilon$. The essential ingredient in all of the above is the fact 
that $M$ is a large decoupled scale much above all the other scales in 
question.

\section{A SUSY example}
\label{bi}

We take here the prototype model for symmetry nonrestoration of Ref. 
\cite{dt96}, which is basically a Wess-Zumino model with a discrete 
symmetry ${\bf D}: \Phi \to -\Phi$ and the addition of a non-renormalizable 
interaction term:

\beq
W = -{1 \over 2} \mu \Phi^2 + {1 \over 4 M} \Phi^4\;,
\label{spot}
\eeq

\noindent
where $M\gg\mu$. This leads to the scalar potential

\bea
V&=& |\phi|^2 |-\mu + {\phi^2 \over M}|^2 \nonumber \\
 &=& {\mu^2 \over 2} (\phi_1^2 + \phi_2^2)  - 
{\mu \over 2M} (\phi_1^4 - \phi_2^4) + \nonumber \\
& &{1 \over 8 M^2} (\phi_1^2 + \phi_2^2)^3\;, 
\label{pot} 
\eea

\noindent
where $\phi = (\phi_1 + i \phi_2)/\sqrt{2}$ is the scalar component of the 
chiral Wess-Zumino superfield $\Phi$. Notice that $\phi_1$ has a negative
 quartic self coupling. 
At $T=0$, as usual, one finds a set of two degenerate minima: 
$\lv\phi\rv = 0$ and $\lv \phi \rv^2 = \mu M$. The usual 1-loop induced
 correction to the
 effective potential at high T is now

\beq 
\Delta V_{\rm 1-loop}(T) = {T^2 \over 8} |
{\partial^2 W \over \partial \phi^2}|^2 =
 {T^2 \over 8} |-\mu + {3 \phi^2 \over M} |^2
\label{Tpot1}
\eeq

\noindent or

\bea
\Delta V_{\rm 1-loop}(T)&=& 
{T^2 \over 8} {\Large [}\mu^2 - 3{\mu\over M} 
(\phi_1^2 - \phi_2^2) \nonumber \\
& &+ {9 \over 4 M^2} (\phi_1^2 + \phi_2^2)^2 {\Large ]}\;.
\label{Tpot2}
\eea

Notice that here the field $\phi_1$ plays precisely the role of the field
 $\phi$ in section \ref{ai}. Thus, we would again conclude (erroneously) 
that the symmetry is not restored at high T. However, as shown in section
 \ref{ai}, at two loops one finds the dominant $(T^4/M^2) \phi^2$ term.

Using the superpotential (\ref{spot}) and the usual rules for the evaluation
 of Feynman diagrams in thermal field theory \cite{w74,dj74}, it
 is straightforward to calculate this contribution as \cite{bms96}

\beq
\Delta V_{\rm 2-loops}(T) = {9 T^4\over 32 M^2 } |\phi|^2 = 
{9 T^4\over 64 M^2} (\phi_1^2 + \phi_2^2)\;.
\label{Tpot3}
\eeq

Obviously, just as in the example of the real field $\phi$ 
in section \ref{ai},
the symmetry $\phi \to -\phi$ gets clearly restored at high temperature.

\section{ General Discussion}
\label{ci}

We would like to address here the issue of the convergence of the perturbation 
theory at high temperature for a general nonrenormalizable theory. Inspired
 by the application to SUSY, we take a case of a complex scalar field in the
 limit of the vanishing $T=0$ mass, and for simplicity we imagine a larger 
$U(1)$ global symmetry. Then the general form of the non-renormalizable
 potential is

\beq
V = \sum_{n=0}^\infty C_{2n} {|\phi|^{2 n +4} \over M^{2 n}}\;.
\eeq

Now, of course, one has to go to all the loops and it is not a hard 
exercise to obtain the following leading $T-$dependent mass for $\phi$
 (along the lines of Ref. \cite{bms96})

\beq
m^2_T = \sum_{n=0}^\infty C_{2n} {(n+2) (n+2)! \over M^{2 n}} \left(
{T^2 \over 12}\right)^{n+1}\;.
\label{mti}
\eeq

For large orders of perturbation theory (at least if $C_{2n}$ is not 
vanishing fast enough), one has the usual factorial growth which
 indicates the nonconvergence of the perturbation theory. This is reminiscent
 of all the known field theories (see e.g the recent review with references
 \cite{f95}). However. in some cases, such as the case in which the $C_{2n}$
 alternate the signs and do not grow too fast, the perturbation theory can be
 given a meaning through the prescription of Borel. 

Borel's prescription says the following. If the series 

\beq
f(z) = \sum_{n=1}^\infty a_n z^n
\label{s3}
\eeq

\noindent diverges, try instead

\beq
f(z) = \int_0^\infty db e^{-b/z} \tilde f(b)\;,
\label{s4}
\eeq

\noindent where

\beq
 \tilde f(b) = \sum_{n=0}^\infty a_{n+1} {b^n \over n!}
\label{s5}
\eeq

\noindent is called the Borel transform of $f(z)$. Of course, when $f(z)$
 in (\ref{s3}) converges, the expressions (\ref{s3}) and (\ref{s4})
 are completely equivalent.

To make this more transparent, we take a simple example

\beq
V = { |\phi|^6 /M^2 \over 1 + |\phi|^2/M^2}\;.
\label{s6}
\eeq

This is the potential that one obtains at the tree-level in the
 renormalizable version of the SUSY example of section \ref{ai} 
with $\mu=0$:

\beq
W = M X^2 + X \phi^2\;.
\label{s7}
\eeq

After integrating out the heavy field $X$, one obtains the effective 
potential in (\ref{s6}).

In this case we have

\bea
C_0 &=& 0 \nonumber \\
C_{2 n}&=& (-1)^{n+1} \label{s8}  \; \; , \; \; n>0
\eea

\noindent and the Borel prescription gives 

\beq
m_T^2 = {T^2 \over 12} g(x) \label{s8a} \;,
\eeq

\noindent where

\beq
g(x) = \int_0^\infty db e^{-b/x} {18 - 6 b \over (1 + b)^5}
\label{s9}
\eeq

\noindent and $x = T^2/(12 M^2)$.
Since $x\ll1$, its is easy to see that in (\ref{s9}) most of the contribution
comes from $b$ in the vicinity of the origin, where $18 \gg 6 b$. 

Then
 $g(x) > 0$ and we get a positive leading high $T$ mass term

\beq
m_T^2 > 0 \;.\label{s10}
\eeq

This is to be expected, since it is 
known that the asymptotic series converges
 to the Borel result up to terms $\sim 1/x$. Since $x$ is extremely small, 
the sum converges up to very high orders in perturbation theory. In other
 words, up to an error of order $x^2$, 
the Borel summed result in (\ref{s8a})-(\ref{s9}) is equal
 to the leading two-loop result in section \ref{bi} 
(in order to compare with (\ref{Tpot3}), the result (\ref{s8a}) 
must be multiplied by a factor of $(3/2)^2$, since in (\ref{Tpot3}) 
also the fermion loops were considered). However, it is 
reassuring to know that one can give a meaning to an infinite diverging
 sum for (\ref{mti}).

\section{Summary and Outlook}
\label{sum}

As clear from the above, the internal symmetry in supersymmetric theories 
seem to get restored at high temperature, even in the presence of 
non-renormalizable interactions. However, one must admit that the proof 
offered is valid only for a single chiral superfield. We suspect that 
the above is true in general, but we have not been able to come 
up with an explicit proof. It remains a challenge to do so or 
to find a counterexample \cite{gd}. A potential counterexample 
could result from derivative couplings \cite{gd}. In fact derivative 
couplings ought to be present in any nonrenormalizable field 
theory. In SUSY, even assuming only terms with maximally two 
derivatives, one can show that the bosonic part of a 
nonrenormalizable Lagrangian for one chiral superfield 
takes the form \cite{sw82} 

\beq
{\cal L}={\partial^2 K\over \partial\phi\partial\phi^*}
|\partial\phi|^2-{1\over
{\partial^2 K\over \partial\phi\partial\phi^*}}
\left|{d W\over d \phi}\right|^2 \;,
\label{weinberg}
\eeq

\noindent
where $K=K(\phi,\phi^*)$ is the K\"ahler potential and 
$W=W(\phi)$ is the superpotential. 
For the example of Section \ref{bi}, after integrating out the 
heavy field $X$, one gets from (\ref{s7}) (valid for $\mu=0$)

\bea
{\partial^2 K\over \partial\phi\partial\phi^*}
&=&1+{|\phi|^2\over M^2}\;,\\
W&=&{\phi^4\over 4M}\;,
\eea

\noindent
which from (\ref{weinberg}) gives 

\beq
{\cal L}_{INT}={|\phi|^2|\partial\phi|^2\over M^2}-
{|\phi|^6/M^2\over 1+|\phi|^2/M^2}\;.
\eeq

The second term has been already discussed above, so let 
us now concentrate on the first term. It contributes to the 
1-loop mass correction as (for $\mu\ne 0$) 

\bea
\Delta m_T^2&=&{i\over M^2}\int_T {d^4 k\over (2\pi)^4} 
{k^2\over k^2-\mu^2}\nonumber\\
&=&-{T\over M^2}\sum_n\int {d^3 \vec{k}\over (2\pi)^3} 
{(2\pi n T)^2+\vec{k}^2\over (2\pi n T)^2+\vec{k}^2+\mu^2}
\nonumber\;,\\
\eea

\noindent
which is potentially of the same order $T^4/M^2$, that 
we kept before. However, it can be shown after some 
thought that in the limit $\mu=0$ (which is our leading 
approximation) it vanishes.

If it were true that internal symmetries in SUSY are always 
restored, one would remain with the cosmological problem of 
domain walls and monopoles. Of course, inflation remains 
as a possible way out, however it must take place after the 
creation of these topological defects. In the case of domain 
walls, it may be that the quantum gravitational effects 
break discrete symmetries and thus provide a way of preventing 
these objects from dominating the energy density of the 
universe \cite{rs94}.

\section*{Acknowledgements}

We are deeply grateful to Alejandra Melfo for her help in 
preparing this manuscript and for the collaboration on most 
of the work presented here. We also thank Gia Dvali for his 
deep interest and his comments and suggestions. G.S. wishes 
to thank Rabi Mohapatra and other organizers of SUSY96 for 
the excellent job in organizing this stimulating meeting, 
and Francesco Vissani for the days of Planet X. B.B. wishes to 
thank the hospitality of the ICTP High Energy Group and the 
Ministry of Science and Technology of Slovenia for partial 
support.


\begin{thebibliography}{999999}
\bibitem{gd} Gia Dvali keeps coming to us with potential counterexamples 
             and it may be a question of time before one of them works.
\end{thebibliography}
\end{document}